\documentstyle[12pt,a4wide]{article}
\newcommand{\msbarscheme}{\overline{\mbox{\scriptsize MS}}}

\begin{document}

\thispagestyle{empty}
\begin{flushright}
MZ-TH/97-03\\
hep-ph/9703268\\
March 1997\\
\end{flushright}
\vspace{0.5cm}
\begin{center}
{\Large\bf Resummation analysis of the $\tau$ decay width}\\[.3truecm]
{\Large\bf using the four-loop $\beta$-function}\\[1truecm]
{\large S.~Groote,$^1$ J.G.~K\"orner$^1$ 
and A.A.~Pivovarov$^{1,2}$}\\[.7cm]
$^1$ Institut f\"ur Physik, Johannes-Gutenberg-Universit\"at,\\[.2truecm]
  Staudinger Weg 7, D-55099 Mainz, Germany\\[.5truecm]
$^2$ Institute for Nuclear Research of the\\[.2truecm]
  Russian Academy of Sciences, Moscow 117312
\vspace{1truecm}
\end{center}

\begin{abstract}
We extract the strong coupling constant $\alpha_s(m_\tau^2)$ from the 
semileptonic $\tau$ decay width taking into account resummation effects 
from the running of the strong coupling constant. In the $\overline{\rm MS}$ 
scheme the result reads $\alpha_s=0.375\pm 0.007$ to third order and 
$\alpha_s=0.378\pm 0.007$ to fourth order in the $\beta$-function, 
respectively, where we use the recently computed four-loop coefficient 
$\beta_3$. These values for the coupling constant have to be compared with 
the value $\alpha_s=0.354\pm 0.005$ derived from a third order 
analysis of $\tau$ decays. We determine the exact value of the convergence 
radius of the perturbation series by analyzing the singularity structure of 
the complex coupling constant plane.
\end{abstract}

\newpage

The precise determination of the strength of the strong interactions is of 
paramount importance for phenomenological applications of the Standard 
Model. Within perturbative QCD the strength of the strong interaction is 
quantitatively measured by the coupling constant $\alpha_s$. Although the 
value of $\alpha_s$ depends on the choice of renormalization scheme in 
finite orders of perturbation theory, the running coupling constant 
$\alpha_s(Q^2)$ has now become a widely recognized measure of the strength 
of the strong interactions~\cite{altrev}.

It has been realized since many years that $\tau$ decays provide an ideal 
setting for obtaining accurate information on strong interactions in the low 
energy domain. The inclusive semileptonic $\tau$ decay width was analyzed in 
detail within perturbative QCD including also the small nonperturbative 
power suppressed effects. A numerical value of $\alpha_s$ was obtained from 
this analysis~\cite{Braaten}. The error that was quoted in these analyses, 
however, accounts mainly for the statistical uncertainty in the experimental 
decay width $\Gamma_\tau$ and does not include systematic errors due to the 
truncation of the perturbation series itself. This problem has widely been 
discussed in the literature and many possible solutions have been proposed. 
Suggestions for the improvement of the analysis include the resummation of 
$\pi^2$ corrections using analytical continuation~\cite{tau} (see 
also~\cite{Pich}), the analysis of scheme dependence~\cite{Chyla}, and the 
inclusion of possible infrared renormalon contributions~\cite{Altarelli}.

In this letter we present an analysis along the lines of~\cite{tau,Pich} 
which consists of a partial resummation of the perturbation series for the 
semileptonic $\tau$ width. We take advantage of the new result on the 
fourth order coefficient $c_3$ of the beta function~\cite{Ritbergen} in the 
series expansion
\begin{equation}\label{betafunction}
\beta(a)=-a^2(1+ca+c_2a^2+c_3a^3+\ldots)
\end{equation}
where $a=\beta_0\alpha_s/4\pi=9\alpha_s/4\pi$ for $N_c=3$ and $n_f=3$. 
The specific choice of normalization of the coupling constant $a$ used in
Eq.~(\ref{betafunction}) is convenient in higher order applications of the 
renormalization group since it removes all factors $\beta_0$ from the 
relevant perturbative formulas. As an additional bonus the magnitude of all 
numbers become of order unity in the perturbative expansion. The connection 
with the traditional normalization is $c_i=\beta_i/(\beta_0)^{i+1}$. 

The expression for the semileptonic $\tau$ decay ratio has the 
form~\cite{Braaten}
\begin{equation}\label{tautheor}
R_\tau=\frac{\Gamma(\tau\rightarrow\nu_\tau+{\rm hadrons})}
  {\Gamma(\tau\rightarrow\nu_\tau+\mu^-+\bar\nu_\mu)}
  =\int_0^{m_\tau^2}W_\tau\left(\frac{s}{m_\tau^2}\right)
  \frac{\rho(s)ds}{m_\tau^2}
\end{equation}
with the weight factor
\begin{equation}\label{weight}
W_\tau(x)=2(1-x)^2(1+2x).
\end{equation}
The spectral density $\rho(s)={\sl Im\,}\Pi(s)/\pi$ is defined in terms of 
the correlator $\Pi(Q^2)$ for charged semileptonic currents (see
e.g.~\cite{tau,Pich}).

We define a reduced $\tau$ decay ratio $r_\tau$ by writing the semileptonic 
$\tau$ decay ratio in a factorized form as
\begin{equation}\label{Rtau}
R_\tau=R_\tau^0\Big(1+\frac{4}{9}r_\tau\Big).
\end{equation}
$R_\tau^0$ is the lowest order partonic value of the ratio $R_\tau$ which 
is given by the number of colours $N_c$ to lowest order in the electroweak 
interaction. Using the experimental value 
$R_\tau^{\rm exp}=3.649\pm0.014$~\cite{Pich2} and $R_\tau^0=3$ one 
calculates
\begin{equation}\label{rtauexp}
r_\tau^{\rm exp}=0.487\pm 0.011.
\end{equation}
It is well known that the reduced ratio $r_\tau$ in Eq.~(\ref{Rtau}) has 
the power series expansion 
\begin{equation}
r_\tau=a(m_\tau^2)+\dots.
\end{equation}

We stress that corrections due to power suppressed terms, electroweak 
interactions and other possible corrections (mass terms, for instance)
can easily be taken into account by a readjustment of $R_\tau^0$ leaving the 
analysis of the hadronic part $r_\tau$ of the perturbation series untouched. 
A readjustment of $R_\tau^0$ naturally implies a readjustment of the 
numerical value of $r_\tau^{\rm exp}$ which is the only input parameter for 
our numerical estimates in massless QCD with $n_f=3$. 

We briefly outline the basic notions of the resummation analysis of the 
perturbative series of $r_\tau$. The analysis is done within the 
$\overline{\rm MS}$ scheme. Needless to say the resummation analysis can of 
course be performed in any other given scheme. In the resummation analysis 
one resums all those terms in the perturbation series that are accessible 
through the perturbative solution of the renormalization group 
equation~\cite{tau,Pich}.

To start with, we define a reduced Adler's function $d(Q^2)$ in analogy to 
Eq.~(\ref{Rtau}) by writing
\begin{equation}
D(Q^2)=\frac1{4\pi^2}\Big(1+\frac49d(Q^2)\Big)
\end{equation}
for Adler's function $D(Q^2)$. In the Euclidean domain this reduced function 
has the expansion
\begin{equation}
\label{dfunc}
d(Q^2)=a(Q^2)+k_1a^2(Q^2)+k_2a^3(Q^2)+k_3a^4(Q^2)+\ldots
\end{equation}
The reduced rate function $r_\tau$ can then be expressed with the help of 
the $\beta$-function coefficients $c$, $c_i$ in Eq.~(\ref{betafunction}) and 
the coefficients $k_i$ in Eq.~(\ref{dfunc}). One has~\cite{Braaten}

\begin{eqnarray}\label{gen}
r_\tau&=&a_\tau\Bigg[1+\frac{19}{12}a_\tau+\left(\frac{265}{72}
  +\frac{19}{12}c-\frac{\pi^2}3\right)a_\tau^2
  \nonumber\\&&\qquad\qquad\qquad
  +\left(\frac{3355}{288}+\frac{1325}{144}c+\frac{19}{12}c_2
  -\frac{19}{12}\pi^2-\frac56c\pi^2\right)a_\tau^3+\ldots\Bigg]\nonumber\\&&
  +k_1a_\tau^2\Bigg[1+\frac{19}6a_\tau+\left(\frac{265}{24}+\frac{19}{6}c
  -\pi^2\right)a_\tau^2+\ldots\Bigg]\nonumber\\&&
  +k_2a_\tau^3\Bigg[1+\frac{19}4a_\tau+\ldots\Bigg]
  +k_3a_\tau^4\Big[1+\ldots\Big]
\end{eqnarray}
where $a_\tau=a(m_\tau^2)$. In the expansion Eq.~(\ref{gen}) we have 
explicitly listed terms up to and including $a_\tau^4$. Up to $O(a_\tau^3)$ 
all coefficients are known. For completeness we have also listed the 
$a_\tau^4$ contribution with its unknown $k_3$-piece. Numerically one has
\begin{equation}\label{num}
r_\tau=a_\tau+2.312a_\tau^2+5.208a_\tau^3+(6.848+k_3)a_\tau^4
\end{equation}
where we have omitted the ellipses still present in Eq.~(\ref{gen}). For 
the numerical evaluation in Eq.~(\ref{num}) we have used (see 
e.g.~\cite{ChetKuehn})
\begin{equation}
c=\frac{64}{81}\approx 0.790\quad\mbox{and}\quad
c_2=\frac{3863}{4374}\approx 0.883
\end{equation}
as well as
\begin{equation}
k_1=\frac{299}{54}-4\zeta(3)\approx 0.729\qquad\mbox{and}\qquad
k_2=\frac{58057}{1458}-\frac{3116}{81}\zeta(3)+\frac{200}{27}\zeta(5)
  \approx 1.258
\end{equation}
The five-loop coefficient $k_3$ is not yet known. Later on we present a 
simple estimate of $k_3$ which is used to obtain a feeling for the 
sensitivity of our results on the value of $k_3$. It is noteworthy that 
the recently calculated four-loop coefficient~\cite{Ritbergen} of the 
$\beta$-function,
\begin{equation}
c_3=\frac{140599}{118098}+\frac{3560}{6561}\zeta(3)\approx 1.843
\end{equation}
does not show up yet in this fixed order of perturbation theory (up to 
$O(a_\tau^4)$). The coefficient $c_3$ does, however, enter in our 
resummation analysis. Note that the series in the square brackets in each 
line of Eq.~(\ref{gen}) depend only on the running of the coupling constant 
determined by the $\beta$-function and the weight function $W_\tau$ in 
the integral~(\ref{tautheor}). Therefore the corresponding ``horizontal'' 
series are accessible to resummation. This has been done before up to the 
third order in the $\beta$-function~\cite{tau,Pich}. In the present paper 
we go beyond the analysis in~\cite{tau,Pich} by including the recently 
calculated four-loop coefficient of the $\beta$-function~\cite{Ritbergen}.

The results of resumming the series in Eq.~(\ref{gen}) can in general be 
presented in the form
\begin{equation}\label{genform}
r_\tau=a_\tau M_0(a_\tau;\beta)+k_1a_\tau^2M_1(a_\tau;\beta)
  +k_2a_\tau^3M_2(a_\tau;\beta)+k_3a_\tau^4M_3(a_\tau;\beta)+\ldots 
\end{equation}
where the expansion coefficients $M_i(a_\tau;\beta)$ have no implicit 
dependence on the coefficients $k_i$ and reflect the effect of the running 
of the coupling constant and the dependence on the specific weight function 
$W_\tau(x)$ in Eq.~(\ref{tautheor}). Explicit fixed order $O(a_\tau^4)$ 
expansions of these coefficients are given in Eq.~(\ref{gen}). The 
coefficient functions $M_i(a_\tau;\beta)$ are completely determined by the 
$\beta$-function and can be calculated explicitly using the renormalization 
group equation for the running coupling constant $a$ with the starting value 
$a_\tau$. Thus the coefficient functions can be obtained to all orders in 
the coupling constant $a_\tau$ in resummed form using the appropriate 
approximation for the $\beta$-function. We emphasize that after resummation 
the formal ordering in terms of powers of the coupling constant $a_\tau$ in 
Eq.~(\ref{gen}) is lost (compare~\cite{OPT}).

Let us return to the integral representation Eq.~(\ref{tautheor}) which we 
shall rewrite in reduced form according to Eq.~(\ref{Rtau}). Accordingly we 
define a reduced part of the correlator by $p(Q^2)$ and integrate $p(Q^2)$ 
along the contour $|s|=m_\tau^2$ in the complex $Q^2=-s$ plane. We obtain
\begin{equation}\label{defin}
r_\tau=\int_0^{m_\tau^2}W_\tau\left(\frac{s}{m_\tau^2}\right)
  \frac{r(s)ds}{m_\tau^2}
  =-\frac1{2\pi i}\oint_{|s|=m_\tau^2}W_\tau\left(\frac{s}{m_\tau^2}\right)
  \frac{p(s)ds}{m_\tau^2}
\end{equation}
The function $p(-Q^2)$ is calculated by integrating the differential equation
\begin{equation}
d(Q^2)=-Q^2\frac{dp(-Q^2)}{dQ^2}.
\end{equation}

Some words of caution are in order concerning the use of Eq.~(\ref{defin}). 
In the second part of Eq.~(\ref{defin}) we have assumed that one can use 
the perturbative approximation for $p(Q^2)$ along the contour $|s|=m_\tau^2$. 
Due to the existence of Landau singularities (corresponding to poles even in 
lowest order of perturbation theory) the perturbative expression for $p(s)$ 
does not have the correct analyticity properties as prescribed by the 
K\"all\'en--Lehmann representation of the full correlator function. The 
result of doing the contour integral therefore also depends on whether the 
contour encircles the Landau singularity or not. An alternative definition 
of the integral on the left hand side of Eq.~(\ref{defin}) would be to use 
the discontinuity of the renormalization group improved correlator function 
$p(Q^2)$ across the positive semi-axis. These two definitions of the 
integral of the renormalization group improved spectral density differ by 
an amount $(\Lambda^2/m_\tau^2)^n$, where $\Lambda$ is the usual scale 
of QCD and $n$ is determined by the power of $x$ in the weight function 
Eq.~(\ref{weight}). In perturbation theory one has a pure polynomial 
expansion in terms of the coupling constant. As a result of analytically 
continuing the polynomial terms in the complex plane one obtains factors of 
$(\ln z)^n$. They all have the correct analyticity properties (cut along 
the positive semi-axis) and the above ambiguity does not occur in any 
finite order of perturbation theory. Redefinitions of this kind (or direct 
treatment of running on the positive semi-axis) were considered in the 
literature~\cite{Ste,KrasPiv,Grunfix}. We do not further dwell upon this 
point but consider the right hand side of Eq.~(\ref{defin}) as a definition 
of the left hand side keeping in mind the above possibility of an 
alternative definition.

In order to proceed with the resummation of the perturbation series (for 
early references see e.g.~\cite{anal}) we decompose the coefficient 
functions $M_i(a_\tau;\beta)$ into moment functions according to 
\begin{equation}\label{mom}
M_{i,n}(a_\tau;\beta)=\frac{n+1}{2\pi i}\oint_{|x|=1}(-x)^na^i(m_\tau^2x)dx.
\end{equation}
Considering the weight function $W_\tau$ in Eq.~(\ref{weight}) one has
\begin{equation}\label{lincom}
M_i(a_\tau;\beta)=2M_{i,0}(a_\tau;\beta)-2M_{i,2}(a_\tau;\beta)
  +M_{i,3}(a_\tau;\beta).  
\end{equation}
These moments constitute the building blocks of our analysis.

As an instructive example we consider the resummation of the moment 
function $M_{1,0}(a_\tau;\beta)$ to leading order in the $\beta$-function, 
i.e.\ $\beta(a)=-a^2$. To this order of the $\beta$-function, the moment 
$M_{1,0}(a_\tau;\beta)$ corresponds to the $a^2(Q^2)$-term in $d(Q^2)$ 
which in turn is contributed to by the $a(Q^2)$-term in $p(Q^2)$ such that
\begin{equation}\label{example}
a_\tau^2M_{1,0}(a_\tau;\beta)=-\frac1{2\pi i}\oint_{|x|=1}
  a(-m_\tau^2x)dx=\frac1{2\pi}\int_{-\pi}^\pi
  \frac{a_\tau e^{i\phi}d\phi}{1+ia_\tau\phi}.
\end{equation}
One finally has
\begin{equation}
M_{1,0}(a_\tau;\beta)={1\over 2\pi a_\tau}\int_{-\pi}^\pi
  \frac{e^{i\phi}d\phi}{1+ia_\tau\phi},
\end{equation}
where $a_\tau=a(m_\tau^2)=(\ln(m_\tau^2/\Lambda^2))^{-1}$ is the solution 
of the renormalization group equation to lowest order. The generalization 
of this technique to higher orders in the $\beta$-function and the reduced 
Adler's function $d(Q^2)$ in Eq.~(\ref{dfunc}) is quite 
straightforward~\cite{tau,Pich} and will not be described in much detail. 
We have extensively used MATHEMATICA~\cite{Math} for the fourth order 
analysis and found it to be a very appropriate tool for doing these 
calculations.

In this way we can determine values of the strong coupling constant using 
the resummed form of Eq.~(\ref{gen}) as written down in Eq.~(\ref{genform}). 
In order to exhibit the dependence of the resulting strong coupling constant 
on the order of the $\beta$-function accuracy we list the resummed result 
for increasing orders of $\beta$-function accuracy. We obtain
\begin{eqnarray}\label{resres}
a_\tau^{(1)}&=&0.2666\pm 0.0041,\qquad
a_\tau^{(2)}\ =\ 0.2668\pm 0.0045,\nonumber\\
a_\tau^{(3)}&=&0.2687\pm 0.0048,\qquad
a_\tau^{(4)}\ =\ 0.2704\pm 0.0050
\end{eqnarray}
or, in terms of $\alpha_{\msbarscheme}$,
\begin{eqnarray}\label{alpres}
\alpha_\tau^{(1)}&=&0.3723\pm 0.0058,\qquad
\alpha_\tau^{(2)}\ =\ 0.3726\pm 0.0063,\nonumber\\ 
\alpha_\tau^{(3)}&=&0.3751\pm 0.0067,\qquad
\alpha_\tau^{(4)}\ =\ 0.3775\pm 0.0071
\end{eqnarray}
where $a_\tau^{(1)}$ refers to the use of the first order $\beta$-function 
$\beta(a)=-a^2$, etc. For the sake of comparison we also give the result of 
calculating $a_\tau$ to fixed order in perturbation theory using 
Eq.~(\ref{num}) to third order in $a_\tau$. One obtains 
$a_\tau^{(3)}=0.2535\pm 0.0035$ (i.e.\ $\alpha_\tau^{(3)}=0.3540\pm 0.0048$). 
When we quote errors on these numbers the error is entirely determined by the 
experimental error on $r_\tau^{\rm exp}$ as given in Eq.~(\ref{rtauexp}).

We postpone the discussion of the numerical results in Eqs.~(\ref{resres}) 
and~(\ref{alpres}) until after analyzing the convergence properties of the 
resummed functions. The first attempt to determine the convergence radius of 
the resummed functions has been done in~\cite{Pich}. The authors 
of~\cite{Pich} obtained approximate values for the convergence radius by 
numerical investigation. Here we present a new method to determine the 
convergence radius by analyzing the singularity structure in the complex 
$a_\tau$-plane.

In the simple lowest order example given above by Eq.~(\ref{example}), the 
radius of convergence is determined by the solution of the equation 
$1-i\pi a=0$, leading to the region of convergence $|a|<1/\pi$~\cite{Pich}. 
In higher orders of $\beta$-function accuracy the investigation of the 
convergence properties of the resummed functions $M_{i,n}(a_\tau,\beta)$ in 
Eq.~(\ref{mom}) is not as simple. It is clear that the numerical 
investigation employed by the authors of~\cite{Pich} is no longer efficient 
as one gets closer and closer to the true radius of convergence. In fact 
one can do better than that. The key observation is that we can search for 
convergence radii as singularities of the resummed function 
$M_{i,n}(a_\tau,\beta)$ continuing $a_\tau$ to the whole complex plane. The 
evolution of $a$ along the contour $Q^2=m_\tau^2e^{i\phi}$ 
($\phi\in[-\pi,\pi]$) is governed by the renormalization group equation
\begin{equation}\label{RGE}
-i\frac{\partial a}{\partial\phi}=\beta(a)=-a^2(1+ca+c_2a^2+c_3a^3+\ldots\ )
\end{equation}
with any given complex $a_\tau$ as starting value. When $a_\tau$ is real, it 
is clear that there is no singularity of the solution of the renormalization 
group equation~(\ref{RGE}) and thus no singularity of the resummed functions 
$M_{i,n}(a_\tau,\beta)$. In fact, the search for singularity solutions 
of Eq.~(\ref{RGE}) has to be extended to the whole complex $a_\tau$-plane. 
The closest singularity in the complex $a_\tau$-plane then determines the 
convergence radius of the resummed functions $M_{i,n}(a_\tau,\beta)$.

Technically one searches for the singular solutions of the renormalization 
group equation~(\ref{RGE}) by analyzing the solution to~(\ref{RGE}) in 
implicit form,
\begin{equation}\label{RGEsol}
i\phi(a;a_\tau)=\int^a_{a_\tau}\frac{da'}{\beta(a')}.
\end{equation}
The integral is a line integral in the complex $a$-plane connecting the 
complex starting value $a_\tau$ with its evolved value $a$. 
Eq.~(\ref{RGEsol}) determines the value of the angle $\phi$ in the 
evolution from $a_\tau$ to $a$ with $\phi(a=a_\tau;a_\tau)=0$. 
Singularities in the evolution process are encountered at angles 
$\phi_s(a_\tau)=\phi(a=\infty;a_\tau)$. After inversion the singular 
starting values $a_\tau(\phi_s)$ appear as lines in the complex 
$a_\tau$-plane which are parametrized by the angles $\phi_s$ at which the 
singularity $a=\infty$ occurs. The search for the singularity lines in the 
complex $a_\tau$-plane can be conducted with the help of built-in 
facilities of MATHEMATICA. For example, in Fig.~1 we have drawn the 
singularity lines $a_\tau^{(4)}(\phi_s)$ in the $a_\tau$-plane for 
four-loop $\beta$-function accuracy. The singularity closest to the origin 
determines the radius of convergence $\bar a_\tau^{(4)}$ of the resummed 
series $M_{i,n}(a_\tau)$ to four-loop $\beta$-function accuracy.

Let us now list the values of the radii of convergence for increasing 
orders of $\beta$-function accuracy including the one-loop result. One has
\begin{eqnarray}\label{radii}
\bar a_\tau^{(1)}=1/\pi=0.318&\qquad&
  (\bar\alpha^{(1)}=0.444),\nonumber\\
\bar a_\tau^{(2)}=0.744/\pi=0.237&\qquad&
  (\bar\alpha_\tau^{(2)}=0.331),\nonumber\\
\bar a_\tau^{(3)}=0.697/\pi=0.222&\qquad&
  (\bar\alpha_\tau^{(3)}=0.310),\nonumber\\
\bar a_\tau^{(4)}=0.674/\pi=0.214&\qquad&
  (\bar\alpha_\tau^{(4)}=0.299),
\end{eqnarray}
where the quoted numbers for $\bar a_\tau^{(2)}$, $\bar a_\tau^{(3)}$ and 
$\bar a_\tau^{(4)}$ have been calculated up to three-digit accuracy. We 
mention that the algorithmic procedure allows one to determine the radii 
of convergence to any desired degree of accuracy. We have added the 
corresponding numbers for the conventional coupling constant 
$\alpha_\tau=\alpha_s(m_\tau^2)=4\pi a_\tau/9$ in brackets. The authors 
of~\cite{Pich} have determined an upper bound for the three-loop convergence 
radius $\bar a_\tau^{(3)}$. They quote $\bar a_\tau^{(3)}<0.25$ which is 
$12\%$ above the true convergence radius $\bar a_\tau^{(3)}$ listed in 
Eq.~(\ref{radii}).

Looking at the numbers in Eq.~(\ref{radii}) one sees that the convergence 
radii become smaller as the order of the $\beta$-function accuracy 
increases. With the newly calculated four-loop $\beta$-coefficient $c_3$ 
the convergence radius is reduced by $3.6\%$ relative to the three-loop 
result. The $3.6\%$ reduction in the convergence radius must be contrasted 
with the mere $0.6\%$ increase in the value of $a_\tau$ when going from 
three-loop to four-loop order in the $\beta$-function (see 
Eq.~(\ref{radii})). It is interesting to speculate that the shrinking of 
the convergence radius continues as one goes to ever higher orders of 
$\beta$-function accuracy including the possibility that the convergence 
radius shrinks to zero when the order of the perturbative $\beta$-function 
expansion goes to infinity.

As mentioned after Eq.~(\ref{alpres}), the fixed third order value of 
$a_\tau$ is given by $a_\tau^{(3)}=0.2535$ which clearly is outside the 
convergence region. This means that the perturbative approximation for 
moments diverges at the scale determined by the experimental data for the 
semileptonic $\tau$ decay width. The resummed results for $a_\tau$ in 
Eq.~(\ref{resres}) lie outside the respective convergence regions. The 
message is clear. The resummed values are not accessible by using higher 
and higher order approximations of perturbation theory.

Let us illustrate the lack of convergence for the fixed order perturbation 
series of $r_\tau$ using this value $a_\tau^{(3)}=0.2535$. In 
Table~\ref{tab1} we list the resulting values of $r_\tau$ for increasing 
orders in $a_\tau$ for this case.
\begin{table}\begin{center}
\begin{tabular}{|r|l||r|l||r|l|}\hline
 $n$&$r_\tau$&$n$&$r_\tau$&$n$&$r_\tau$\\\hline
 $1$&$0.2535$&$5$&$0.4993$&$9$&$0.7400$\\
 $2$&$0.4021$&$6$&$0.4622$&$10$&$1.5577$\\
 $3$&$0.4870$&$7$&$0.4402$&$11$&$4.059$\\
 $4$&$0.5153$&$8$&$0.4894$&$12$&$11.905$\\\hline
\end{tabular}
\caption{Higher order perturbative predictions for the reduced part of the 
semileptonic $\tau$ decay width using $a_\tau^{(3)}=0.2535$\label{tab1}}
\end{center}\end{table}
The $n=3$ value for $r_\tau$ is fixed by the experimental input value 
$r_\tau^{\rm exp}=0.487$ by construction. Nothing dramatic happens to the 
series up to $n=8$. Starting with $n=9$ the series can be seen to show a 
divergent behaviour.

Let us now return to discuss the resummed coupling values in 
Eq.~(\ref{resres}). Increasing the $\beta$-function accuracy does not 
appear to change the numerical results much. In order to exhibit some of 
the structure of our resummation analysis we present results on the moment 
functions $M_{i,n}(\alpha;\beta)$ and their relevant linear combination
Eq.~(\ref{lincom}) in Tables~\ref{tab2} and~\ref{tab3}.
\begin{table}\begin{center}
\begin{tabular}{|r|l|l|l|l|l|}\hline
$i$&$r_\tau$&$M_i(a_\tau;\beta)$&$M_{i,0}(a_\tau;\beta)$
  &$M_{i,2}(a_\tau;\beta)$&$M_{i,3}(a_\tau;\beta)$\\\hline
$0$&$0.371693$&$1.383455$&$1.111405$&$0.823075$&$0.806795$\\
$1$&$0.080542$&$1.530973$&$1.050332$&$0.546269$&$0.522848$\\
$2$&$0.034765$&$1.424447$&$0.880180$&$0.299298$&$0.262683$\\\hline
$\Sigma$&$0.487000$&&&&\\\hline
\end{tabular}
\caption{Resummed moments for $\beta(a)=-a^2(1+ca+c_2a^2)$\label{tab2}}
\end{center}\end{table}
\begin{table}\begin{center}
\begin{tabular}{|r|l|l|l|l|l|}\hline
$i$&$r_\tau$&$M_i(a_\tau;\beta)$&$M_{i,0}(a_\tau;\beta)$
  &$M_{i,2}(a_\tau;\beta)$&$M_{i,3}(a_\tau;\beta)$\\\hline
$0$&$0.372502$&$1.377756$&$1.105213$&$0.815529$&$0.798389$\\
$1$&$0.080255$&$1.506411$&$1.035635$&$0.537942$&$0.511026$\\
$2$&$0.034243$&$1.376758$&$0.859464$&$0.297142$&$0.252113$\\\hline
$\Sigma$&$0.487000$&&&&\\\hline
\end{tabular}
\caption{Resummed moments for $\beta(a)=-a^2(1+ca+c_2a^2+c_3a^3)$\label{tab3}}
\end{center}\end{table}
In Table~\ref{tab2} we have used three-loop $\beta$-function accuracy 
relevant for the determination of the resummed value of $a_\tau^{(3)}$ in 
Eq.~(\ref{resres}) while Table~\ref{tab3} contains our result on the 
resummed value of $a_\tau^{(4)}$ up to four-loop $\beta$-function accuracy. 
For the sake of comparison we list in Table~\ref{tab4} values for the fixed 
third order expansion for the corresponding moments.
\begin{table}\begin{center}
\begin{tabular}{|r|l|l|l|l|l|}\hline
$i$&$r_\tau$&$M_i(a_\tau;\beta)$&$M_{i,0}(a_\tau;\beta)$
  &$M_{i,2}(a_\tau;\beta)$&$M_{i,3}(a_\tau;\beta)$\\\hline
$0$&$0.38204$&$1.50693$&$1.22140$&$0.90427$&$0.87266$\\
$1$&$0.08445$&$1.80282$&$1.50705$&$1.16902$&$1.12676$\\
$2$&$0.02051$&$1.00000$&$1.00000$&$1.00000$&$1.00000$\\\hline
$\Sigma$&$0.48700$&&&&\\\hline
\end{tabular}
\caption{Moments for the fixed third order expansion\label{tab4}}
\end{center}\end{table}
Comparing the numbers in Tables~\ref{tab2} and~\ref{tab3} one sees a weak 
dependence on which order of the $\beta$-function is used. The dependence 
is sowewhat larger for the coefficients $i=1$ and $i=2$, but the 
contributions of these coefficients to the total result are suppressed. 
The bulk contribution comes from the leading order term ($i=0$). Comparing 
Tables~\ref{tab2} and~\ref{tab3} on the one hand and Table~\ref{tab4} on 
the other hand one sees that the effect of resummation is largest for the 
higher order coefficients $i=1,2$. However, the effect of the higher order 
coefficients on the total result is small. The overall effect of 
resummation at this level is not significant. Resummation does, however, 
lead to a slight increase of the value of the strong coupling constant 
extracted from $\tau$ decay data though, namely from $\alpha_s=0.3540$ to 
$\alpha_s=0.3775$. Nevertheless, such an increase must be taken seriously 
considering the level of accuracy that the $\alpha_s$ determination has 
reached in other experiments. We shall return to this point later on.

The value of $k_3$ is still unknown. One can obtain a first estimate for 
the value of $k_3$ by means of a Pad\'e approximation (e.g.~\cite{Ellis}). 
Using this estimate one can get a hint on the sensitivity of our results on 
higher order corrections in the $\overline{\rm MS}$ scheme. The naive 
$(1,1)$-Pad\'e approximation for the function $d(Q^2)$ results in the value 
$k_3=k_2^2/k_1=2.17$~\cite{Pich,KatSta}. Performing the resummation 
analysis with this value of $k_3$ one finds a value 
$a_\tau^{(4)}(k_3=k_2^2/k_1)=0.2650$ 
($\alpha_\tau^{(4)}(k_3=k_2^2/k_1)=0.3700$).

We want to emphasize that the renormalization group equation and therefore 
our resummation procedure is invariant under a one-parameter
subgroup parametrized by $\gamma$ which describes the rescaling of the 
renormalization scale $m_\tau$. The coupling parameters in the two schemes 
are related by
\begin{equation}
a'(m_\tau^2)=a(e^\gamma m_\tau^2)
  =a-\gamma a^2+(\gamma^2-c\gamma)a^3
  -(\gamma^3-\frac52c\gamma^2+c_2\gamma)a^4+\ldots.
\end{equation}
Rewriting the reduced $d(Q^2)$ function in terms of $a'$ leads to a 
rescaling of the coefficients $k_i$ but leaves the $\beta$-function 
invariant. A suitable choice of the parameter $\gamma$, which corresponds to 
a different choice of renormalization scheme, may lead to a value $a'$ for 
the coupling {\em inside\/} the convergence region. At the same time one may 
have deteriorated the series in $k_i$. This possibility underlines the fact 
that the coupling constant in a particular scheme has no physical meaning.

To get rid of this ambiguity, we choose a physical observable, namely the 
reduced Adler's function itself as an effective coupling 
constant~\cite{Grunberg,Kscheme},
\begin{equation}
d(Q^2)=a_k(Q^2).
\end{equation}
For this effective coupling constant the $\beta$-function is given by 
\begin{equation}
\beta_k(a_k)=-a_k^2(1+\rho_1a_k+\rho_2a_k^2+\rho_3a_k^3)
\end{equation}
with
\begin{eqnarray}
  \rho_1&=&c,\qquad
  \rho_2\ =\ c_2-ck_1-k_1^2+k_2,\nonumber\\
  \rho_3&=&c_3-2c_2k_1+ck_1^2+4k_1^3-6k_1k_2+2k_3.
\end{eqnarray}
Because the $\beta$-function is expressed in terms of a physical observable, 
the coefficients $\rho_1$, $\rho_2$ and $\rho_3$ are renormalization group 
invariants~\cite{Steinv} (see e.g.~\cite{Maxwell}). Now there is no gain in 
knowing $c_3$ in the $\overline{\rm MS}$ scheme because the four-loop 
coefficient $c_3$ of the $\beta$-function appears together with the unknown 
coefficient $k_3$ in the invariant $\rho_3$. This observation reflects the 
fact that one needs full information on the $c_i$ and $k_i$ for a given 
physical observable in any given renormalization scheme. Numerically we 
obtain $\rho_1\approx 0.790$, $\rho_2\approx 1.035$ and
\begin{eqnarray}
\label{rho3}
\rho_3&=&-\frac{37096148}{59049}+\frac{4820288}{6561}\zeta(3)
  +\frac{12352}{81}\zeta(3)^2-256\zeta(3)^3
  -\frac{59800}{243}\zeta(5)\nonumber\\&&\qquad\qquad
  +\frac{1600}9\zeta(3)\zeta(5)+2k_3\ =\ -2.97953+2k_3\approx 1.360
\end{eqnarray}
using the Pad\'e-approximation $k_3=k_2^2/k_1=2.17$. The region of 
convergence in this scheme can be calculated to be $|a_k^{(4)}|<0.214$ and 
again does not include the value $a_k^{(4)}=0.3630$.
 
From the explicit results in Eq.~(\ref{rho3}) we also see that there is an 
interesting possibility of having the infrared fixed point in the Euclidean 
region if the coefficient $k_3$ happens to be small enough. In naive 
perturbation theory the sign of $k_3$ is not very important (as well as in 
the resummation approach within the $\overline{\rm MS}$ scheme) while in 
the effective coupling scheme things can change qualitatively for negative 
$\rho_3$. Because physics is determined by invariant quantities this would 
mean that the scheme dependence stops to be soft in this order of 
perturbation theory.

Let us summarize our main numerical results. We have analyzed semileptonic 
$\tau$ decay data using a partially resummed perturbation series and have 
obtained a value of the strong coupling constant which is approximately 
$7\%$ higher than the value obtained from the fixed third order analysis. 
In the resummed analysis we find $\alpha_{\msbarscheme}=0.3775$ compared to 
$\alpha_{\msbarscheme}=0.3540$ in the fixed order expansion analysis. When 
we rescale the results up to the $Z$ mass, the resummed value of 
$\alpha_{\msbarscheme}$ from $\tau$ decay tends to be in better accord with 
the determination of $\alpha_s$ from $Z$ decay data~\cite{altrev}. This is 
good. On the other hand a higher value of $\alpha_s$ is not welcome when 
one does low energy sum rule phenomenology~\cite{Shifman}. Even the present 
fixed order value $\alpha_s=0.3540$ is rather large when one wants to 
justify the perturbative computation of coefficient functions in the 
framework of the operator product expansion.

Judging from the results presented in this paper there does not seem to be 
much room for improvements on the accuracy of our results. Adding the new 
four-loop term to the $\beta$-function has not changed the results of the 
resummation analysis by a significant amount ($a_\tau^{(3)}$ compared to 
$a_\tau^{(4)}$ in Eq.~(\ref{resres})). Apart from the fact that the five-loop 
result is difficult to come by we do not anticipate that adding the 
five-loop term to the $\beta$-function will affect our resummation results 
much. We have attempted to estimate the influence of the five-loop $k_3$ 
term. The analysis has shown that higher order $k_i$-contributions tend to 
be numerically suppressed. A pattern of such a suppression can be seen in 
Eq.~(\ref{num}) with $k_3=k_2^2/k_1=2.17$. Although this is not a 
renormalization group invariant statement, there is the temptation to skip 
the higher order $k_i$ contributions altogether and use the resummed 
remainder as an improvement of perturbation theory.

It is very likely that our approach makes more sense than to push fixed 
order perturbation theory to ever higher orders. For example, the fixed 
fifth order formulas that have been written down in~\cite{KatSta} within 
the latter approach may not even lead to an improved analysis. As our 
numerical results have shown, the resummed values of $\alpha_s$ from $\tau$ 
decay lie outside the convergence radii and can therefore not be obtained 
from a power series expansion. Regular perturbation series do not converge 
to their resummed counterparts. The experimental value of $r_\tau$ appears 
to be too large for a fixed order perturbation analysis to apply.

We end our discussion with some critical remarks. The simple analyticity 
properties of the two-point correlator (K\"all\'en--Lehmann representation) 
describing semileptonic $\tau$ decays are central to the present resummation 
analysis. In other more complicated cases a corresponding resummation 
analysis may not even exist. We do not even attempt to answer the question 
whether it makes sense to compare two values of $\alpha_s$ derived from 
different experiments, where the one value is extracted using a resummation 
analysis and the other value is obtained using fixed order perturbation 
theory.

Last but not least, we reiterate that the entire scheme presented here is 
based on a particular prescription of analytic continuation in the complex 
$Q^2$-plane which corresponds to a particular prescription for treating 
the Landau pole inside the integration contour. If one chooses another 
prescription, the derived values for $\alpha_{\msbarscheme}$ will change. 
This uncertainty is inherent to the perturbative solution of the 
renormalization group equation since the perturbative solution has wrong 
analyticity properties in the complex $Q^2$-plane.\\[1truecm]
\noindent{\bf Acknowledgements:}
This work is partially supported by the BMBF, FRG, under contract 
No.~06MZ865. A.A.P.\ greatfully acknowledges partial financial supported 
by the Russian Fund for Basic Research under contracts No. 96-01-01860
and 97-02-17065 and by INTAS-93-0744. The work of S.G.\ is supported by 
the DFG, FRG.

\vspace{1truecm}

\centerline{\Large\bf Figure Caption}
\vspace{.5cm}
\newcounter{fig}
\begin{list}{\bf\rm Fig.\ \arabic{fig}:}{\usecounter{fig}
\labelwidth1.6cm\leftmargin2.5cm\labelsep.4cm\itemsep0ex plus.2ex}
\item Singular starting values $a_\tau^{(4)}(\phi_s)$ 
($\phi_s\in[-\pi,\pi]$) as parametric curves in the complex plane. The radius 
of the dashed circle determines the convergence radius of the resummed 
series. The resummed value $a_\tau^{(4)}=0.2704$ extracted from the $\tau$ 
decay rate lies outside the region of convergence.
\end{list}
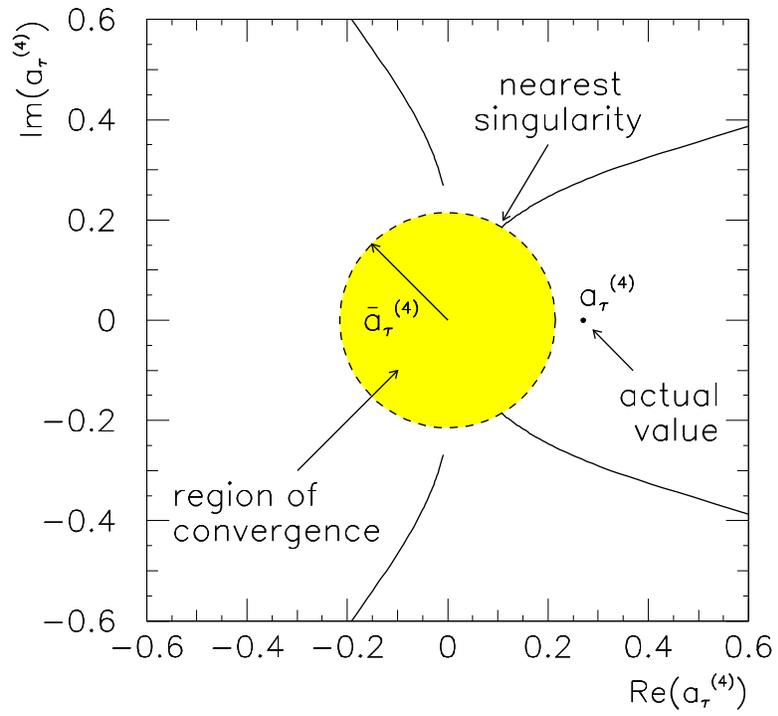
\end{document}